\documentclass[12pt]{article}

\usepackage{graphicx}
\usepackage{amssymb}
\usepackage{amsmath}

\numberwithin{equation}{section}

\begin{document}
\baselineskip=16pt
\begin{titlepage}

\flushright{
MISC-2011-12}

\begin{center}
\vspace*{12mm}

{\Large\bf%
One loop divergences in quantum
gravity coupled with nonlocal matter fields
}\vspace*{10mm}

Takayuki Hirayama
\vspace*{4mm}

{\it Maskawa Institute for Science and Culture, 
Kyoto Sangyo University, 
\\
Kyoto, 603-8555, Japan
}\\[1mm]

\end{center}
\vspace*{10mm}

\begin{abstract}\noindent%
We study the matter one-loop quantum corrections to the gravitational sector in a gravity theory coupled with a nonlocal scalar field. We find that non-renormalizable divergences disappear when the propagator of the scalar field approaches to a non-zero constant at high momentum, while the vacuum energy is still divergent. We also comment on
non-local modification of fermion, gauge field and graviton.


\end{abstract}

\end{titlepage}

\newpage


\section{Introduction}

The quantum gravity, by quantizing general relativity around the flat space by treating the metric fluctuations around the flat metric as quantum fluctuations, is known to be non-renormalizable. 't Hooft and Veltman computed the one loop quantum corrections of scalar fields give rise to non-renormalizable divergences to the gravity sector~\cite{'tHooft:1974bx}. Similarly the one loop quantum corrections of fermion and gauge fields also introduce non-renormalizable divergences~\cite{Deser:1974cz}. Sagnotti and Ven showed the two loop graviton graphs introduce non-renormalizable divergences~\cite{Goroff:1985sz}. The non-renormalizability is a serious problem in quantum gravity and this leads us the deep thought on the microphysics of quantum gravity.

Stelle showed that a higher derivative gravity $\!\int\!\! \sqrt{-g} \{\kappa^{-2}R +\alpha R_{\mu\nu}R^{\mu\nu} -\beta R^2\}$
 turns out to be renormalizable~\cite{Stelle:1976gc}, since the graviton propagator behaves like $1/p^4$ for a large momentum resulting the suppression of divergences of loop graphs. However the graviton propagator has additional poles other than the one for massless graviton in general $\alpha$ and $\beta$, and the new poles correspond to massive graviton and massive scalar particles. The massive graviton has a wrong kinetic term, i.e. ghost particle and then the unitarity is violated. If one fine tunes the couplings $\alpha$ and $\beta$ such that the mass of massive graviton goes to infinity, the renormalizability is lost. In fact the appearance of ghost particles is a general feature in a higher derivative theory. To avoid the violation of unitarity in a higher derivative theory, Lee and Wick~\cite{Lee:1969fy} proposed that ghost particles can decay into normal particles (the mass of ghost particles acquires a non-zero imaginary part), and then the asymptotic states do not contain ghost particles and the unitarity is not violated. However multi ghost particles will have a real mass and will appear in the asymptotic states. Therefore the idea of Lee and Wick is still questionable.
 
One may try to use an exponentially dumping propagator at high momentum, for example $e^{-p^2/M^2}/(p^2+m^2)$, and expect that a renormalizable quantum gravity can be constructed. The exponentially dumping propagator does not have additional poles other than the one for normal particle and it seems that the problem of the violation of unitarity does not exist. Since the propagator contains infinite space/time derivatives, the theory becomes a non-local theory and a string field theory also has this type of propagator. In a non-local theory, we suspect the violation of causality and unitarity. The causality in string field theory is considered in~\cite{Lowe:1994ns} and the commutator of string fields is in general non-vanishing outside the string light cone. This fact does not necessarily mean that signals can propagate faster than the speed of light, but is merely the result of trying to describe extended objects by local fields. In a large but finite N QCD, extended object glue balls are highly non locally interacting, but the underling theory is QCD and is apparently causal. The violation of causality is computed and grows with energy grows in a non-local $\phi^4$ theory~\cite{Haque:2008ud}. Therefore one should discuss whether a causality is violated in each non-local theory.

The violation of unitarity is also suspected in a non-local theory. Although the Feynman propagator does not contain a ghost like component, it is computed in a quantum theory in a non-commutative spacetime that the perturbative unitarity is violated~\cite{Gomis:2000zz}. Hayashi gave a general procedure of constructing a unitariy S-matrix using Yang-Feldman formalism in a non-local in time theory~\cite{hayashi}. In the case of quantum theory in non-commutative spacetime, one can use a modified time ordering and S-matrix is perturbatively unitary~\cite{Chu:2002fe}. However the propagator defined by using the modified time ordering is no longer the Feynman propagator and thus the positive energy flow, i.e. only the positive (negative) energy particles travel forward (backward) in time, is lost~\cite{Fujikawa:2004rt}. There is another example, a quantum theory in Poincare invariant commutative nonassociative spacetimes, 
where the unitarity and positive energy condition are both satisfied~\cite{Sasai:2006ua}. One should thus check the violation of unitarity in each non local theory.

If the action contains more than second time derivatives, one need more than two initial conditions to specify the solution of equations of motion. This is true for finite time derivatives, but not true for infinite time derivatives and for example we only need two initial conditions when the action is $\phi e^{\partial^2/M^2}(\partial^2+m^2)\phi$\footnote{
$e^{-p^2/M^2}$ has an essential pole at $p^2\rightarrow -\infty$. We may have to specify how to treat this essential pole.
}~\cite{Barnaby:2007ve}.

Although the suspected problems in a non-local theory exist, one might be able to resolve them positively. Therefore it is still important to study whether the non-locality really tame ultra violate divergences and we can define a renormalizable quantum gravity. In this paper we focus on the matter quantum corrections to the gravity sector. 't Hooft and Veltman showed the existence of non-renormalizable infinities in the scalar one loop corrections. We modify the scalar propagator which becomes a non local one (and thus no new poles appear) at high energy and study whether non-renormalizable infinities can disappear. We show that a exponential dumping does not improve the UV behaviour of loop graphs since the gravitational couplings on the other hand are enhanced. Interestingly we find that the propagator should approach to a non-zero constant, and then the non-renormalizable divergences disappear\footnote{
Ho~\cite{Ho:2003zm} studied the renormalization to the Newton constant and obtain the same result. In his computation $\sqrt{-g}$ is not taken into account, but his result does not change.}. 
This is an unexpected result since the UV behaviour of propagator is worsen compared with $1/p^2$. However approaching to a non-zero constant implies that the matter becomes non-dynamical, and it is understandable that if the matter becomes non-dynamical, the matter does not induce non-renormalizable quantum corrections.

We will comment on the same modification of propagator for fermion, gauge and graviton fields. Since the propagator is modified, we have to study again the renormalizability of Yukawa and gauge couplings. In the next section, we will give computation of matter one loop graphs in quantum gravity coupled with a non-local real scalar field. In section 3, we discuss the modification of propagators for fermion, gauge and graviton and restudy the renormalizability of standard model. In section 4, we give our conclusions.


\section{Non local scalar field couples to general relativity}

As described in Introduction, we expect that non-renormalizable UV divergences disappear if we modify the scalar propagator at high momentum. Our action is Euclidean general relativity coupled with a real scalar field,
\begin{align}
 I &= \!\int \! d^4x \sqrt{g} \Big[ \frac{-1}{16\pi G} (R-2\Lambda)
 +\{ \phi f(\nabla^2) \phi + h.c. \} \Big]
 ,
 \\
 Z &= \!\int \! {\cal D}[g]{\cal D}[\phi] \: e^{-I},
\end{align}
where we have ignored the surface terms and the signature is $(++++)$. The matter action if it is canonical is 
$\phi f(\nabla^2) \phi=\phi(-\nabla^2+m^2)\phi$, but we do not specify the form of
$f(\nabla^2)$ here.

We assume the flat space is a classical solution ($\Lambda=0$ and $\phi=0$ at the classical level), and treat the metric fluctuations around the flat metric as quantum fluctuations. We define the path integral which corresponds to a naive quantization and 
a $T^*$ product correlation function can be computed from the path integral~\cite{Fujikawa:2004rt}.

The metric is expanded around the flat metric,
\begin{align}
 g_{ij} = \delta_{ij} +\delta g_{ij},
 \hspace{3ex}
 ( g^{ij} = \delta^{ij} +\delta g^{ij} )
 \label{mf}
\end{align}
and we denote $\delta g_{ij} = \kappa h_{ij}$, $\kappa$ is a dimensionless parameter in this paper (and thus $h_{ij}$ is also dimensionless), and then 
$\delta g^{ij} = -\kappa h^{ij} + \kappa^2 h^{ik}h_{k}^{~j}+\cdots$
and $\sqrt{g}=1+\kappa (h/2) +\kappa^2 (-(h^{ij}h_{ij}/4)+(h^2/8))+\cdots$.
 The Lorentz indices are made up and down by using the background metric $\delta_{ij}$.
We expand $\nabla^2\phi$ in terms of $\kappa$, and we obtain
\begin{align}
 \nabla^2 \phi &= g^{ij}\partial_i\partial_j\phi -g^{ij}\Gamma^k_{~ij}\partial_k\phi
 =\frac{1}{\sqrt{g}}\partial_i ( \sqrt{g} g^{ij} \partial_j \phi)
 = g^{ij}\partial_i\partial_j\phi +(\partial_i g^{ij})\partial_j\phi
 +\frac{1}{2}g^{kl}(\partial_i g_{kl}) g^{ij}\partial_j \phi
 \nonumber\\
 &= \partial^2 \phi + \kappa \Big( - h^{ij}\partial_i\partial_j\phi
 -(\partial_i h^{ij}) \partial_j \phi
 +\frac{1}{2}(\partial_i h) \partial^i\phi \Big)
 \nonumber\\
 &\:\:\:\:
 +\kappa^2 \Big(
 h^{ik}h_{k}^{j}\partial_i\partial_j \phi
 +(\partial_i (h^{ik}h_k^j))\partial_j\phi
 -\frac{1}{4}\partial_i(h_{jk}h^{jk})\eta^{ij}\partial_j\phi
 -\frac{1}{2}(\partial_i h) h^{ij}\partial_j\phi \Big)
 +\cdots
 ,
\end{align}
where $\partial^2=\delta^{ij}\partial_i\partial_j$ etc.
We notice here that since the metric is dimensionless and the matter action is
$\sqrt{g}\phi f(\nabla^2) \phi$, the momentum dimension of one graviton and two scalar vertex read from $\phi \nabla^2\phi$ is two, ie. $\sim \partial_i\partial_j$, which is the same as that of the inverse of the propagator (the classical value of $\sqrt{g}$ equals $1$ ). 
We denote the three point vertex read from $\phi f(\nabla^2) \phi$
as $\tau_{ij}^B(p_g,p_a,p_b)$ where $p_g$, $p_a$ and $p_b$ are the momentum of graviton and two scalars,  Therefore if $f(\nabla^2)$ behaves as $(\nabla^2)^n$ at high momentum, the momentum dimension of one graviton and two scalar vertex, $\sim \partial^{2(n-1)}\partial_i\partial_j$, is the same as that of the inverse of propagator, $1/f(\partial^2)$. Another three point vertex, $\tau^A_{ij}(p_g,p_a,p_b)$, comes from taking the first order in $\kappa$ from $\sqrt{g}$, ie $(h/2)\phi f(\partial^2) \phi$. Here since this vertex is proportional to $f(\partial^2)$, this interaction is enhanced if the propagator is dumped at high momentum.
For $f(\nabla^2)=e^{-\nabla/M^2}$ and thus an exponential dumping propagator, the momentum dimension of $\tau^A$ is $\sim e^{-\partial^2}$ and that of $\tau^B$ is $e^{-\partial^2}\partial_i\partial_j$.

The four point vertices which are two graviton and two scalar vertices are categorized into three by taking the (second, first, 0-th) order in $\kappa$ from $\sqrt{g}$ and the (0-th, first, second) order from $\phi f(\nabla^2) \phi$ respectively. We denote them
$\tau^{A',B',C'}_{ijkl}(p_{g1},p_{g2},p_a,p_b)$. We notice again that if $f(\nabla^2)=(\nabla^2)^n$, the momentum dimension of $\tau_{ijkl}^{A'}$, $\tau_{ijkl}^{B'}$, and $\tau_{ijkl}^{C'}$ vertices are all $\partial^{2n}$ at high momentum because of the dimensionless of the metric.


\subsection{One loop quantum corrections}

Now we compute one loop matter graphs, which are vacuum graphs, tadpole graphs and graviton self energy graphs.

The scalar vacuum graph with loop momentum $l$ is easily computed
\begin{align}
 \int\!\! d^4l \frac{1}{f(-l^2)} f(-l^2) \sim \Lambda_{cut}^4
 \hspace{3ex}\rightarrow
 \delta {\cal L} \sim \sqrt{g}\Lambda_{cut}^4
\end{align} 
where hereafter we ignored order one coefficients such as $1/(2\pi)^4$ since we are only interested in the power of UV divergences. We introduce the UV cut off $\Lambda_{cut}$ to estimate the power of UV divergences and $\delta {\cal L}$ denotes the quantum corrections to the Lagrangian.
We can easily see that the vacuum energy is quartic divergent independent of the form of $f(\nabla^2)$ and so the cosmological constant should be renormalized.
The graviton tadpole one loop matter graphs using $\tau^A_{ij}$ and $\tau^B_{ij}$ are computed
\begin{align}
 &\int\!\! d^4l \frac{1}{f(-l^2)} \tau_{ij}^A(0,l,-l) \sim
 \int\!\! d^4l \frac{1}{f(-l^2)} \delta_{ij}^Af(-l^2) \sim
 \Lambda_{cut}^4
 &&\hspace{3ex}
 \rightarrow 
 \delta {\cal L} \sim \Lambda_{cut}^4 h
 \\
 &\int\!\! d^4l \frac{1}{f(-l^2)} \tau^B_{ij}(0,l,-l) 
 &&\hspace{3ex}\rightarrow
 \delta {\cal L} \sim (h,\partial^2 h,\partial^i\partial^j h_{ij})
 .\label{t2}
\end{align}
It is easily seen that the tadpole graph using $\tau^A$ contributes to the cosmological constant. Since $\tau^B$ picks us the momentum of graviton up to two times, the one loop tadpole graph induces the renormalization to the cosmological constant and the Newton constant in front of $R$.
If $f(\nabla^2)$ at high momentum behaves as $(\nabla^2)$ or $e^{-\nabla^2/M^2}$, the momentum dimension of $\tau^B_{ij}/f(-l^2)$ are zero or two and then the UV divergences of~\eqref{t2} are computed form the dimensional counting,
\begin{align}
 \delta {\cal L} &\sim (\Lambda_{cut}^4 h, \Lambda_{cut}^2 \partial^2 h, \Lambda_{cut}^2\partial^i\partial^j h_{ij} )
 &\rightarrow& \sqrt{g}\Lambda_{cut}^4, \sqrt{g} \Lambda_{cut}^2 R
 &&\mbox{for } (\nabla^2)^n
 \\
 &\sim (\frac{\Lambda_{cut}^6}{M^2} h, \frac{\Lambda_{cut}^4}{M^2} \partial^2 h,   
 \frac{\Lambda_{cut}^4}{M^2}\partial^i\partial^j h_{ij} )
 &\rightarrow& \sqrt{g}\frac{\Lambda_{cut}^6}{M^2}, 
 \sqrt{g} \frac{\Lambda_{cut}^4}{M^2} R
 &&\mbox{for } e^{-\nabla^2/M^2}
\end{align}
and thus we need the renormalization of Newton constant as well. It is interesting that the exponential dumping of propagator on the other hand makes the UV divergences worse. In order to cancel the UV divergences, one understands that the momentum dimension of
the ratio $\tau^B_{ij}/f(-l^2)$ should be smaller than minus four. This is realized if $f(\nabla^2)$ at high momentum approaches to a non-zero constant, for example
\begin{align}
 f(\nabla^2) \rightarrow c_0 + \frac{c_1}{(\nabla^2)^m} 
 + c_2(\nabla^2)^n e^{\nabla^2/M^2}
 \label{cst}
\end{align}
where $c_0\neq0$, $m>3$ and $n$ is arbitrary, since the highest momentum dimension of $f(\nabla^2)$ at high momentum comes from the constant $c_0$ which does not induce the vertex $\tau^B$, and the vertex $\tau^B$ comes from the rest part whose momentum dimension is $1/\partial^{2m}$ when $c_1\neq 0$ and $\partial^{2n+2}e^{\partial^2}$.

The graviton self energy one loop graphs using two $\tau^A$, one $\tau^A$ and one $\tau^B$, and two $\tau^B$ are computed
\begin{align}
 &\int\!\! d^4l \: \frac{\tau^A_{ij}(p,l,-l-p)\tau^A_{kl}(p,l,-l-p)}{f(-l^2)f(-(p-l)^2)}  
 \sim \Lambda_{cut}^4
 &&\hspace{0ex}\rightarrow
 \delta L \sim \Lambda_{cut}^4h^2
 \label{s1}\\
 &\int\!\! d^4l \:\frac{\tau^A_{ij}(p,l,-l-p)\tau^B_{kl}(p,l,-l-p)}{f(-l^2)f(-(p-l)^2)}  
 &&\hspace{0ex}\rightarrow
 \delta L \sim h(h,\partial^2 h,\partial^i\partial^j h_{ij})
 \label{s2}\\
 &\int\!\! d^4l \:\frac{\tau^B_{ij}(p,l,-l-p)\tau^B_{kl}(p,l,-l-p)}{f(-l^2)f(-(p-l)^2)} 
 &&\hspace{0ex}\rightarrow
 \delta L \sim (h^2, h\partial^2h,h\partial^i\partial^j h_{ij},
 \nonumber\\ 
 &&&\hspace{7ex} h\partial^4 h,h\partial^2\partial^i\partial^j h_{ij},h_{ij}\partial^i\partial^j\partial^k\partial^l h_{kl})
 \label{s3}
\end{align}
We notice again that $\tau^B$ picks up the momentum of graviton up to two times.
Thus similarly \eqref{s1} and \eqref{s2} induce the renormalization to the cosmological constant and the Newton constant. The power of divergences are also same which are quartic for the cosmological constant and quadratic for the Newton constant if $f(\nabla^2)\rightarrow (\nabla^2)^n$ at high momentum. And if the propagator approaches to a non-zero constant~\eqref{cst}, the divergence for the Newton constant disappears. The graph~\eqref{s3} induces the renormalization to the higher dimensional term, $R^2$ and $R_{ij}R^{ij}$ and the theory is non-renormalizable if the graph is UV divergent. Thus we like to realize this graph is UV finite, and from the power counting we obtain
\begin{align}
 \delta {\cal L} 
 &\rightarrow
 \sqrt{g} (\ln \Lambda_{cut}) (R^2,R_{ij}R^{ij})
 &&\mbox{for } (\nabla^2)^n
 \\
 &\rightarrow
 \sqrt{g} \frac{\Lambda_{cut}^2}{M^2} (R^2,R_{ij}R^{ij})
 &&\mbox{for } e^{-\nabla^2/M^2}
\end{align}
and UV finite if the propagator again approaches to a non-zero constant~\eqref{cst}.

The graviton self energy one loop graphs using $\tau^{A',B',C'}$ are
\begin{align}
 &\int\!\! d^4l \frac{\tau^{A'}_{ijkl}(p,-p,l,-l)}{f(-l^2)}  \sim \Lambda_{cut}^4
 &&\hspace{0ex}\rightarrow
 \Lambda_{cut}^4 h^2
 \\
 &\int\!\! d^4l \frac{\tau^{B'}_{ijkl}(p,-p,l,-l)}{f(-l^2)}  
 &&\hspace{0ex}\rightarrow
 \delta L \sim h(h, \partial^2 h, \partial^i\partial^j h_{ij})
 \\
 &\int\!\! d^4l \frac{\tau^{C'}_{ijkl}(p,-p,l,-l)}{f(-l^2)}  
 &&\hspace{0ex}\rightarrow
 \delta L \sim (h^2, h\partial^2 h, h\partial^i\partial^j h_{ij},
 \nonumber\\
 &&&\hspace{12ex}
 h\partial^4 h, 
 h\partial^2\partial^i\partial^j h_{ij},h_{ij}\partial^i\partial^j\partial^k\partial^l h_{kl})
\end{align}
and the result is the same, ie. if $f(\nabla^2)$ approaches to a non-zero constant, the divergent contribution to the higher dimensional terms $R^2$ and $R_{ij}R^{ij}$ disappears.

In summary, we conclude that if the propagator approaches to a non-zero constant~\eqref{cst}, the one loop quantum corrections are UV finite except for the corrections to the cosmological constant. Therefore in our context, an exponential dumping in the propagator due to a non-locality does not resolve the UV divergences. If $f(\nabla^2)$ approaches to a constant, it means that the scalar field is no longer dynamical. In this sense, it is easy to understand there is no quantum corrections since the scalar field is non dynamical.

One example of the propagator which has no more than two poles (thus non-local) and approaches to a constant at high momentum is
\begin{align}
 f(\nabla^2) = M^2 \tanh(-\nabla^2/M^2) +m^2.
\end{align}


\section{Non local fermion, gauge and graviton fields}

As described in the previous section, a scalar field should become non-dynamical at high momentum to realize UV finite quantum corrections except for the quantum correction to the cosmological constant. The same argument can apply for other matter fields, fermion and gauge fields. Therefore a fermion propagator should also approach to non-zero constant at high energy. However since the mass term is not allowed for unbroken gauge fields, a gauge field cannot be non-dynamical naively since we can always redefine the gauge fields so that the kinetic term is canonically normalized. (In the case of non-dynamical scalar and fermion, if we redefine the fields so that the kinetic term is canonically normalized, the mass term becomes infinity.) Therefore in order to realize non-dynamical gauge fields at high momentum, we may introduce a charged matter field $\Phi$ which takes non-zero vacuum expectation value only at high energy. For example we prepare the scalar potential
\begin{align}
 \lambda e^{\nabla^2/M^2}\left\{|\Phi^\dag\Phi| -v^2\tanh(-\nabla^2/M^2-1) \right\}^2
\end{align}
where we introduced an exponential dumping coupling constant $\lambda e^{\nabla^2/M^2}$. Without the dumping, the renormalizability of the coupling $\lambda$ will be lost since the matter propagators goes to non-zero constant.
Then we introduce $Z(\nabla^2)$ which goes to 
$e^{\nabla^2/M^2}$ at high momentum
in front of the kinetic term for the gauge fields, $\sqrt{g} Z(\nabla^2) g^{ik}g^{jl}F_{ij}F_{jl}$, and then we can realize non-dynamical gauge fields at high momentum. 

If we apply the same argument for the graviton, we have to realize that the graviton becomes non-dynamical at high momentum and thus, for example, introduce the mass terms for graviton (see review~\cite{Hinterbichler:2011tt}) which is non-zero only at high momentum. A recent attempt to modify UV behavior of graviton for constructing a renormalizable quantum gravity is suggested by Horava~\cite{Horava:2009uw}, although it is difficult to avoid the insensitive to the UV modification in IR physics.

Since the propagators approach to a non-zero constant, we should again study the normalizability of Standard Model. It is clear that Yukawa couplings $y$ and Higgs four point interactions as well as the gauge couplings are not renormalizable. We can modify those couplings by non-local couplings such as $e^{\nabla^2/M^2}y$ and then the renormalizability can be recovered.


\section{Conclusion}

We studied one loop quantum corrections in gravity theory coupled with non local scalar fields. We find that if the scalar propagator approaches to non-zero constant, the UV divergences which give the renormalization to Newton constant and higher dimensional terms $R^2$ and $R^{ij}R_{ij}$ disappear. This implies that the scalar field becomes non-dynamical at high momentum and thus there is no quantum corrections. The same argument can apply for other fields. The fermion propagator then should go to a non-zero constant at high momentum so that the fermion becomes non-dynamical and quantum corrections by fermion loops do not give rise the non-renormalizable divergences. For gauge fields, by realizing the gauge symmetry is broken only at high energy, we can realize the gauge fields becomes non-dynamical at high energy. It is interesting to find a mechanism which makes the graviton becomes non-dynamical at high momentum. Since the propagators are modified to be constant at high momentum, the gauge couplings, Yukawa couplings and four point interactions of Higgs fields are not renormalizable in Standard Model. We can modify those couplings, say $y$ by non-local ones such as $e^{\nabla^2/M^2}y$, and the renormalizability of Standard Model is recovered.

If we can treat a non-local scalar fields as infinitely many point particles, a non-local scalar fields contain infinitely many scalar particles. On the other hand a string in string theory contains infinitely many different spin particles. Therefore it is interesting to construct a non-local field which contains different spin particles. In this sense, the bi-local field proposed by Yukawa~\cite{Yukawa:1950eq} might be interesting to be investigated since the bi-local field contains different spin states.


\subsection*{Acknowledgments}

The author would like to thank all the members in Maskawa Institute. 
This work is supported by Maskawa Institute in Kyoto Sangyo University.

\bigskip\bigskip


\begin{thebibliography}{99}



%
\bibitem{'tHooft:1974bx}
  G.~'t Hooft and M.~J.~G.~Veltman,
  ``One loop divergencies in the theory of gravitation,''
  Annales Poincare Phys.\ Theor.\  A {\bf 20} (1974) 69.


%
\bibitem{Deser:1974cz}
  S.~Deser and P.~van Nieuwenhuizen,
  ``One Loop Divergences of Quantized Einstein-Maxwell Fields,''
  Phys.\ Rev.\  D {\bf 10} (1974) 401;
%
%
  ``Nonrenormalizability of the Quantized Dirac-Einstein System,''
  Phys.\ Rev.\  D {\bf 10} (1974) 411;
%
  ``Nonrenormalizability of the Quantized Einstein-Maxwell System,''
  Phys.\ Rev.\ Lett.\  {\bf 32} (1974) 245.
%
%
  S.~Deser, H.~S.~Tsao and P.~van Nieuwenhuizen,
  ``One Loop Divergences of the Einstein Yang-Mills System,''
  Phys.\ Rev.\  D {\bf 10} (1974) 3337.
  


\bibitem{Goroff:1985sz}
  M.~H.~Goroff and A.~Sagnotti,
  ``Quantum Gravity at Two Loops,''
  Phys.\ Lett.\  B {\bf 160} (1985) 81;
%
%
  ``The Ultraviolet Behavior of Einstein Gravity,''
  Nucl.\ Phys.\  B {\bf 266} (1986) 709.
%
%
  A.~E.~M.~van de Ven,
  ``Two loop quantum gravity,''
  Nucl.\ Phys.\  B {\bf 378} (1992) 309.




\bibitem{Stelle:1976gc}
  K.~S.~Stelle,
  ``Renormalization of Higher Derivative Quantum Gravity,''
  Phys.\ Rev.\  D {\bf 16} (1977) 953.


\bibitem{Lee:1969fy}
  T.~D.~Lee and G.~C.~Wick,
  ``Negative Metric and the Unitarity of the S Matrix,''
  Nucl.\ Phys.\  B {\bf 9} (1969) 209.


\bibitem{Lowe:1994ns}
  D.~A.~Lowe, L.~Susskind and J.~Uglum,
  ``Information spreading in interacting string field theory,''
  Phys.\ Lett.\  B {\bf 327} (1994) 226
  [arXiv:hep-th/9402136].


\bibitem{Haque:2008ud}
  A.~Haque and S.~D.~Joglekar,
  ``Possible Detection of Causality Violation in a Non-local Scalar Model,''
  J.\ Phys.\ A  {\bf 42} (2009) 065401
  [arXiv:0808.3179 [hep-th]].


\bibitem{Gomis:2000zz}
  J.~Gomis and T.~Mehen,
  ``Space-time noncommutative field theories and unitarity,''
  Nucl.\ Phys.\  B {\bf 591} (2000) 265
  [arXiv:hep-th/0005129].
  

%
\bibitem{hayashi}
  C.~Hayashi,
  ``Hamiltonian Formalism in Non-local Field Theories,''
  Prog. Theor. Phys. 10 (1953) 533.


\bibitem{Chu:2002fe}
  C.~S.~Chu, J.~Lukierski and W.~J.~Zakrzewski,
  ``Hermitian analyticity, IR / UV mixing and unitarity of noncommutative field
  theories,''
  Nucl.\ Phys.\  B {\bf 632} (2002) 219
  [arXiv:hep-th/0201144].
  

\bibitem{Fujikawa:2004rt}
  K.~Fujikawa,
  ``Path integral for space-time noncommutative field theory,''
  Phys.\ Rev.\  D {\bf 70} (2004) 085006
  [arXiv:hep-th/0406128].


%
\bibitem{Sasai:2006ua}
  Y.~Sasai and N.~Sasakura,
  ``One-loop unitarity of scalar field theories on Poincare invariant
  commutative nonassociative spacetimes,''
  JHEP {\bf 0609} (2006) 046
  [arXiv:hep-th/0604194].


\bibitem{Barnaby:2007ve}
  N.~Barnaby and N.~Kamran,
  ``Dynamics with infinitely many derivatives: The Initial value problem,''
  JHEP {\bf 0802} (2008) 008
  [arXiv:0709.3968 [hep-th]].


\bibitem{Ho:2003zm}
  P.~M.~Ho,
  ``Regularization of Newton constant, transPlanckian dispersion relation, and
  symmetry of particle spectrum,''
  Class.\ Quant.\ Grav.\  {\bf 21} (2004) 2641
  [arXiv:hep-th/0308103].
  

%
\bibitem{Hinterbichler:2011tt}
  K.~Hinterbichler,
  ``Theoretical Aspects of Massive Gravity,''
  arXiv:1105.3735 [hep-th].
  
  
%
\bibitem{Horava:2009uw}
  P.~Horava,
  ``Quantum Gravity at a Lifshitz Point,''
  Phys.\ Rev.\  D {\bf 79} (2009) 084008
  [arXiv:0901.3775 [hep-th]].


%
\bibitem{Yukawa:1950eq}
  H.~Yukawa,
  ``Quantum Theory of Nonlocal Fields. 1. Free Fields,''
  Phys.\ Rev.\  {\bf 77} (1950) 219,
%
  ``Quantum theory of nonlocal fields. 2: Irreducible fields and their interaction,''
  Phys.\ Rev.\  {\bf 80} (1950) 1047,
  ``Structure and mass spectrum of elementary particles. 1: general considerations,''
  Phys.\ Rev.\  {\bf 91} (1953) 415,
  ``Structure and mass spectrum of elementary particles. 2: oscillator model,''
  Phys.\ Rev.\  {\bf 91} (1953) 416.

%




  


\end{thebibliography}
\end{document}